\def\be{\begin{equation}}
\def\ee{\end{equation}}
\def\ba{\begin{array}}
\def\ea{\end{array}}

\def\dps{\displaystyle}

\documentclass[aps,amsmath,amssymb,amsfonts,showpacs,twocolumn,pra]{revtex4}
\usepackage{graphicx}
\def\qed{\leavevmode\unskip\penalty9999 \hbox{}\nobreak\hfill
     \quad\hbox{\leavevmode  \hbox to.77778em{%
               \hfil\vrule   \vbox to.675em%
               {\hrule width.6em\vfil\hrule}\vrule\hfil}}
     \par\vskip3pt}

\newtheorem{theorem}{Theorem}

\begin{document}
\title{Entanglement detection using mutually unbiased measurements}
\author{Bin Chen$^{1}$}
\author{Teng Ma$^{1}$}
\author{Shao-Ming Fei$^{1,2}$}

\affiliation{$^1$School of Mathematical Sciences, Capital Normal University,
Beijing 100048, China\\
$^2$Max-Planck-Institute for Mathematics in the Sciences, 04103
Leipzig, Germany}

\begin{abstract}

We study the entanglement detection by using mutually unbiased
measurements and provide a quantum separability criterion that can
be experimentally implemented for arbitrary
$d$-dimensional bipartite systems. We show that this criterion is more
effective than the criterion based on mutually unbiased bases.
For isotropic states our criterion becomes both necessary and sufficient.

\end{abstract}

\maketitle

\section{Introduction}
Entanglement is one of the most appealing features in the quantum
world and has been extensively investigated in the past decades
\cite{s1,g09}. Being useful resources, quantum entangled states play
key roles in many quantum information prossing tasks, such as
quantum cryptography \cite{s2}, quantum teleportation \cite{s3}, and
dense coding \cite{s4}. An important problem proposed is how to
distinguish quantum entangled states from the states without
entanglement, i.e., separable states. For bipartite pure states,
Schmidt decomposition tells us that a state is separable if and only
if it is of Schmidt rank one. But for mixed states, the problem
becomes formidably complicated. There have been various separability
criteria such as positive partial transposition criterion
\cite{p96,ho96,ho97}, realignment criterion \cite{r1,r2,r3,r4,r5},
covariance matrix criterion \cite{c1}, and correlation matrix
criterion \cite{c2,c3}. Recently, Li \emph{et al}. \cite{li14}
present a generalized form of the correlation matrix criterion for
both bipartite and multipartite quantum systems, which is more
effective than the previous criteria.

Although numerous mathematical tools have been employed in
entanglement detection of given known quantum states,
experimental implementation of entanglement detection for unknown quantum states
has fewer results \cite{Bell,yusxia,liming,zmj}. In Ref. \cite{spe12}, the
authors connected the separability criteria to mutually
unbiased bases (MUBs) \cite{s5} in two-qudit, multipartite and
continuous-variable quantum systems. Based on the correlation
functions, the criterion employs local
measurements only, and can be implemented experimentally. For two-qudit
systems, the criterion is shown to be very powerful
in detecting entanglement of particular states. If $d$ is a prime power
the criterion is both necessary and sufficient for the separability of
isotropic states \cite{be}. However, when $d$ is not a prime power,
the criterion becomes less effective. Generally the
applications of MUBs are subject to the maximum number $N(d)$ of
MUBs. It has been shown that $N(d)$ is no more than $d+1$, and
$N(d)=d+1$ when $d$ is a prime power \cite{s5}. But when $d$ is a
composite number, $N(d)$ is still unknown. Even for $d=6$, we do not
know whether or not there exist four MUBs \cite{s6,s7,s8,s9}.

Recently, Kalev and Gour generalize the concept of MUBs to mutually
unbiased measurements (MUMs) \cite{ka14}. These measurements,
containing  the complete set of MUBs as a special case, need not to
be rank one projectors. Unlike the existence of MUBs which depends
on the dimension of the system, there always exists a complete set
of $d+1$ MUMs, and can be explicitly constructed. MUMs have also
many useful applications in quantum information processing, such as
quantum state tomography \cite{s5,qst1,qst2} and entropic
uncertainty relation \cite{iv92,weh10}.

In this paper, we
study the separability problem by using mutually unbiased
measurements.  We provide a separability criterion for two-qudit
systems. We show that this
criterion is necessary and sufficient for the separability of all
isotropic states in any dimensions.

\section{MUBs and MUMs}
Let us first review some basic definitions of mutually unbiased
bases. Two orthonormal bases
$\mathcal{B}_{1}=\{|b_{i}\rangle\}_{i=1}^{d}$ and
$\mathcal{B}_{2}=\{|c_{j}\rangle\}_{j=1}^{d}$ of $\mathbb{C}^{d}$
are said to be mutually unbiased if
$$
|\langle b_{i}|c_{j}\rangle|=\frac{1}{\sqrt{d}} ,~~~\forall\, i,j=1,2,\cdots,d.
$$
A set of orthonormal bases $\{\mathcal{B}_{1},
\mathcal{B}_{2},\cdots,\mathcal{B}_{m}\}$ in $\mathbb{C}^{d}$ is
called a set of mutually unbiased bases if every pair of bases in
the set are mutually unbiased. Since each $\mathcal{B}_{k}$ can be
written as $d$ rank one projectors summing to  the identity
operator, these MUBs describe $m$ POVM (projective) measurements on a quantum
system of dimension $d$. Thus, if a physical system is prepared in
an eigenstate of basis $\mathcal{B}_{k}$ and measured in basis
$\mathcal{B}_{k'}$, then all the measurement outcomes are equally probable.

Let $\mathcal{B}_{k}=\{|i_{k}\rangle\}_{i=1}^{d},k=1,2,\cdots,m$, be any $m$ MUBs. For any two-qudit state $\rho$, define
$$
I_{m}(\rho)=\sum_{k=1}^{m}\sum_{i=1}^{d}\langle i_{k}|\otimes\langle i_{k}|\rho|i_{k}\rangle\otimes|i_{k}\rangle.
$$
It has been shown that if $\rho$ is separable, then $I_{m}(\rho)\leq1+\frac{m-1}{d}$
\cite{spe12}. For isotropic states
$$
\rho_{iso}=\alpha|\Phi^{+}\rangle\langle\Phi^{+}|+\frac{1-\alpha}{d^{2}}I,
$$
where
$$
|\Phi^{+}\rangle=\frac{1}{\sqrt{d}}\sum_{i=1}^{d}|ii\rangle,~~~ 0\leq\alpha\leq1,
$$
one gets $I_{m}(\rho)=m(\alpha+\frac{1-\alpha}{d})$. Therefore if
$\alpha>\frac{1}{m}$, then $\rho_{iso}$ is entangled by the
criterion. Thus, the power of the criterion depends on $m$. When $d$ is
a prime power, then $m=d+1$, the criterion becomes both necessary and
sufficient for the separability of $\rho_{iso}$,
as $\rho_{iso}$ is entangled for $\alpha>\frac{1}{d+1}$, and
separable for $\alpha\leq\frac{1}{d+1}$ \cite{ber05}.

In Ref. \cite{ka14}, the authors introduced the concept of mutually
unbiased measurements (MUMs). Two POVM  measurements on $\mathbb{C}^{d}$,
$\mathcal{P}^{(b)}=\{P_{n}^{(b)}\}_{n=1}^{d}$, $b=1,2$, are said to be
mutually unbiased measurements if
\begin{equation}
\begin{split}
\mathrm{Tr}(P_{n}^{(b)})&=1,\\
\mathrm{Tr}(P_{n}^{(b)}P_{n'}^{(b')})&=\frac{1}{d},~~~b\neq b',\\
\mathrm{Tr}(P_{n}^{(b)}P_{n'}^{(b)})&=\delta_{n,n'}\,\kappa+(1-\delta_{n,n'})\frac{1-\kappa}{d-1},
\end{split}
\end{equation}
where $\frac{1}{d}<\kappa\leq1$, and $\kappa=1$ if and only if all
$P_{n}^{(b)}$s are rank one projectors, i.e., $\mathcal{P}^{(1)}$
and $\mathcal{P}^{(2)}$ are given by MUBs.

A general construction of $d+1$ MUMs has been presented in \cite{ka14}.
Let $\{F_{n,b}:n=1,2,\cdots,d-1,b=1,2,\cdots,d+1\}$ be a set of
$d^{2}-1$ Hermitian, traceless operators acting on $\mathbb{C}^{d}$,
satisfying $\mathrm{Tr}(F_{n,b}F_{n',b'})=\delta_{n,n'}\delta_{b,b'}$.
Define $d(d+1)$ operators
\begin{equation}
F_{n}^{(b)}=
\begin{cases}
   F^{(b)}-(d+\sqrt{d})F_{n,b},&n=1,2,\cdots,d-1;\\[2mm]
   (1+\sqrt{d})F^{(b)},&n=d,
\end{cases}
\end{equation}
where $F^{(b)}=\sum_{n=1}^{d-1}F_{n,b}$, $b=1,2,\cdots,d+1.$ Then one can construct $d+1$ MUMs explicitly \cite{ka14}:
\begin{equation}\label{3}
P_{n}^{(b)}=\frac{1}{d}I+tF_{n}^{(b)},
\end{equation}
with $b=1,2,\cdots,d+1,n=1,2,\cdots,d,$ and $t$ should be chosen
such that $P_{n}^{(b)}\geq0$. Any $d+1$ MUMs can be expressed in such form.

Corresponding to the construction of MUMs (\ref{3}), the parameter $\kappa$ is given by
$$
\kappa=\frac{1}{d}+t^{2}(1+\sqrt{d})^{2}(d-1).
$$
If the operators $F_{n,b}$ are chosen to be the generalized Gell-Mann operator basis,
one can choose $\kappa=\frac{1}{d}+\frac{2}{d^{2}}$
optimally \cite{ka14}. That is to say, there always exist $d+1$ MUMs
for arbitrary $d$, in contrast to MUBs for which this is possible for prime power $d$.
However, if one fixes the parameter $\kappa$, then the two values of $t$ cannot guarantee the
existence of operator basis such that $P_{n}^{(b)}$s are positive.
For instance, suppose $\kappa=1$ and there exist $d+1$ MUMs, then
it must be a complete set of MUBs. But one does not know whether or not
there exist $d+1$ MUBs when $d$ is not a prime power.

\section{MUMs based separability criterion}
We now generalize the results in Ref. \cite{spe12}, and present a
new separability criterion for two-qudit states by using mutually
unbiased measurements.

\begin{theorem}
Let $\rho$ be a density matrix in $\mathbb{C}^{d}\bigotimes\mathbb{C}^{d}$,
$\{\mathcal{P}^{(b)}\}_{b=1}^{d+1}$  and
$\{\mathcal{Q}^{(b)}\}_{b=1}^{d+1}$ be any two sets of $d+1$ MUMs on
$\mathbb{C}^{d}$ with the same parameter $\kappa$, where
$\mathcal{P}^{(b)}=\{P_{n}^{(b)}\}_{n=1}^{d},\mathcal{Q}^{(b)}=\{Q_{n}^{(b)}\}_{n=1}^{d},b=1,2,\cdots,d+1$.
Define
$J(\rho)=\sum_{b=1}^{d+1}\sum_{n=1}^{d}\mathrm{Tr}(P_{n}^{(b)}\bigotimes
Q_{n}^{(b)}\rho)$. If $\rho$ is separable, then $J(\rho)\leq1+\kappa$.
\end{theorem}

[{\sf Proof}]. We need only to consider pure separable state,
$\rho=|\phi\rangle\langle\phi|\otimes|\psi\rangle\langle\psi|$,
since $J(\rho)$ is a linear function. We have
\begin{eqnarray*}
J(\rho) & = & \sum_{b=1}^{d+1}\sum_{n=1}^{d}\mathrm{Tr}(P_{n}^{(b)}\bigotimes Q_{n}^{(b)}\rho)\\
& = & \sum_{b=1}^{d+1}\sum_{n=1}^{d}\mathrm{Tr}(P_{n}^{(b)}|\phi\rangle\langle\phi|)\mathrm{Tr}(Q_{n}^{(b)}|\psi\rangle\langle\psi|)\\
& \leq &
\frac{1}{2}\sum_{b=1}^{d+1}\sum_{n=1}^{d}\{[\mathrm{Tr}(P_{n}^{(b)}|\phi\rangle\langle\phi|)]^{2}+[\mathrm{Tr}(Q_{n}^{(b)}|\psi\rangle\langle\psi|)]^{2}\}.
\end{eqnarray*}

By using the relation
$$
\sum_{b=1}^{d+1}\sum_{n=1}^{d}(\mathrm{Tr}(P_{n}^{(b)}\rho))^{2}=1+\kappa
$$
for pure state $\rho$ \cite{ka14}, we obtain $J(\rho)\leq1+\kappa$. \quad $\Box$

Note that when $\kappa=1$, our criterion reduces to the previous one
in Ref.\cite{spe12}, which demonstrates that $I_{d+1}(\rho)\leq2$
for all separable states $\rho$,  if there exists a complete set of
MUBs in $\mathbb{C}^{d}$. However, the entanglement detection based
on mutually unbiased measurements is more efficient
than the one based on MUBs  for some states.

Let $\{P_{n}^{(b)}\}_{n=1}^{d}$, $b=1,2,\cdots,d+1,$ be $d+1$ MUMs with
the parameter $\kappa$. Let $\overline{P}_{n}^{(b)}$ denote the
conjugation of $P_{n}^{(b)}$. It is obvious that
$\{\overline{P}_{n}^{(b)}\}_{n=1}^{d}$, $b=1,2,\cdots,d+1,$ is another
$d+1$ MUMs with the same parameter $\kappa$. Then we get
$$
\ba{rcl}
J(\rho_{iso})&=&\dps
\sum_{b=1}^{d+1}\sum_{n=1}^{d}\mathrm{Tr}(P_{n}^{(b)}\bigotimes
\overline{P}_{n}^{(b)}\rho_{iso})\\[4mm]
&=&\dps (d+1)(\alpha\kappa+\frac{1-\alpha}{d}).
\ea
$$
If $\alpha>\frac{1}{d+1}$, then $J(\rho_{iso})>1+\kappa$, and
$\rho_{iso}$ must be entangled by the theorem. Therefore, this
criterion is both necessary and sufficient for the separability of
the isotropic states, namely, it can detect all the entanglement of the isotropic
states (see FIG. 1). It should be emphasized that, unlike the previous
criterion based on MUBs, our criterion works perfectly for any dimension $d$.

\begin{figure}[htpb]
\centering
\includegraphics[width=7cm]{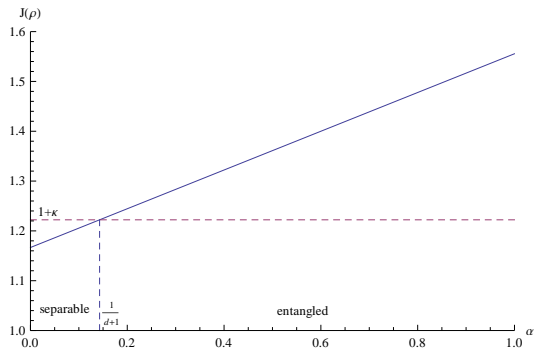}
\caption{\small For isotropic states
$\rho_{iso}$, $J(\rho_{iso})>1+\kappa$ if $\alpha>\frac{1}{d+1}$,
and $\rho_{iso}$ is entangled. This
result coincides with the fact that $\rho_{iso}$ is entangled for
$\alpha>\frac{1}{d+1}$, and separable for $\alpha\leq\frac{1}{d+1}$
\cite{ber05}.} \label{detect1}
\end{figure}

\begin{figure}[htpb]
\centering
\includegraphics[width=7cm]{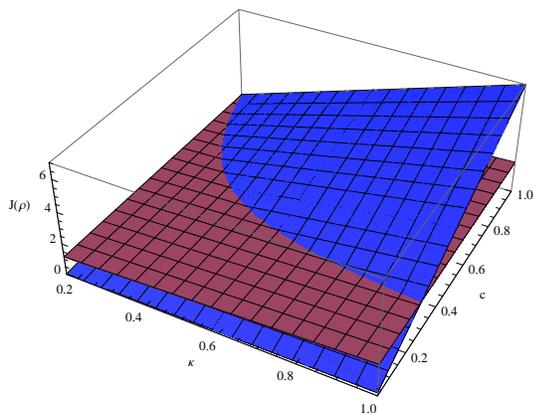}
\caption{\small For Bell-diagonal states, $J(\rho_{Bell})$ violates
the criterion when $c>(1+\frac{1}{\kappa})/(d+1)$. The
criterion detects more entanglement as $\kappa$ increases.}
\label{detect2}
\end{figure}

For a complete set of mutually unbiased measurements, the value of
the parameter $\kappa$ is of great importance. Just as
noted in Ref. \cite{ka14}, for a given choice of operator basis, the
maximal value of $\kappa$ suggests how close we can find a complete
set of MUBs. Therefore, $\kappa$ is expected to be as close as to 1 as
possible.

The efficiency of entanglement detection also depends on the parameter $\kappa$.
Consider the Bell-diagonal states,
\begin{equation*}
    \rho_{Bell}=\sum_{s,t=0}^{d-1}p_{s,t}|\Phi_{s,t}^{+}\rangle\langle\Phi_{s,t}^{+}|,
\end{equation*}
where $p_{s,t}\geq 0$, $\sum_{s,t=0}^{d-1}p_{s,t}=1$, $|\Phi_{s,t}^{+}\rangle=(U_{s,t}\bigotimes
I)|\Phi^{+}\rangle,$
$U_{s,t}=\sum_{j=0}^{d-1}\zeta_{d}^{sj}|j\rangle\langle j\oplus
t|$, $s,t=0,1,\cdots,d-1,$ are Weyl operators,
$\zeta_{d}=e^{\frac{2\pi\sqrt{-1}}{d}}$ and $j\oplus t$ denotes
$(j+t)$ mod $d$. For a proper choice of
$\{\mathcal{P}^{(b)}\}_{b=1}^{d+1}$ and
$\{\mathcal{Q}^{(b)}\}_{b=1}^{d+1}$ with the same parameter
$\kappa$, it is straightforward to obtain
$$
J(\rho_{Bell})\geq c\kappa(d+1),
$$
where $c=\mathrm{max}\{p_{s,t}:s,t=0,1,\cdots,d-1\}$, $\frac{1}{d^{2}}\leq
c\leq 1$. Thus if $c>(1+\frac{1}{\kappa})/(d+1)$, then
$J(\rho_{Bell})>1+\kappa$ and $\rho_{Bell}$ must be entangled by our
criterion. It is obvious that in order to detect more entanglement of
Bell-diagonal states, the parameter $\kappa$ should be as
large as possible (see FIG. 2).

It is interesting to see the connection between the
separability criteria presented in \cite{li14} (and the references therein) and
ours. In fact, given $d+1$ mutually unbiased measurements
$\mathcal{P}^{(b)}=\{P_{n}^{(b)}=\frac{1}{d}I+tF_{n}^{(b)}:~n=1,2,\cdots
d\}$, $b=1,2,\cdots,d+1$, with the parameter
$\kappa=\frac{1}{d}+t^{2}(1+\sqrt{d})^{2}(d-1)$, expanding a
two-qudit state $\rho$ in the operator basis adopted in
$\{\mathcal{P}^{(b)}\}_{b=1}^{d+1}$, we have
$$\ba{rcl}
J(\rho)&=&\dps\sum_{b=1}^{d+1}\sum_{n=1}^{d}\mathrm{Tr}(P_{n}^{(b)}\bigotimes
P_{n}^{(b)}\rho)\\[5mm]
&=&\dps\frac{d+1}{d}+\frac{2(d\kappa-1)}{d-1}\mathrm{Tr}(T),
\ea
$$
where $T$ is the correlation matrix of $\rho$. Thus, if $\rho$ is
separable, then we have $\mathrm{Tr}(T)\leq\frac{d-1}{2d}$, which is just
a special case of the inequality satisfied by separable states in Ref. \cite{li14}.

However, for a general expression
$J(\rho)=\sum_{b=1}^{d+1}\sum_{n=1}^{d}\mathrm{Tr}(P_{n}^{(b)}\bigotimes
Q_{n}^{(b)}\rho)$, we do not know whether or not our criterion can
be deduced to some inequalities contained in Ref.\cite{li14}, since
we do not know the connections between two complete sets of mutually
unbiased measurements with the same parameter $\kappa$. Here we would like to
point out that, unlike the criterion presented in Ref. \cite{li14},
our criterion is based on mutually unbiased measurements, which provides an experimental way
of entanglement detection.

\section{Conclusion and discussions}
We have studied the separability problem via mutually unbiased
measurements. We have presented a new separability criterion for the
separability of two-qudit
states. It has been shown that the criterion based on mutually unbiased
measurements is more efficient than the one based on mutually
unbiased bases. For isotropic states, this criterion is both necessary
and sufficient. It detects all the entangled isotropic states
of arbitrary dimension $d$. The powerfulness of our criterion
is due to that there always exists a complete
set of mutually unbiased measurements, which is not the case for
mutually unbiased bases when $d$ is not a prime power.

Measurement based quantum separability criteria are of
practically significance, as it provides experimental implementation
in detecting entanglement of unknown quantum states. It may be also
interesting to generalize our results to multipartite systems.

\bigskip
\noindent{\bf Acknowledgments}\, \,
This work is supported by the NSFC under number 11275131.


\begin{thebibliography}{18}
\bibitem{s1} R. Horodecki, P. Horodecki, M. Horodecki, and K. Horodecki, Rev. Mod. Phys. \textbf{81}, 865 (2009).

\bibitem{g09} O. Guhne, G. Toth, Phys. Rep. \textbf{474}, 1 (2009).

\bibitem{s2} A. K. Ekert, Phys. Rev. Lett. \textbf{67}, 661 (1991); D. Deutsch, A. Ekert, R. Jozas, C. Macchiavello,
S. Popescu and A. Sanpera, \emph{ibid}. \textbf{77}, 2818 (1996); C. A. Fuchs, N. Gisin, R. B.
Griffiths, C. S. Niu and A. Peres, Phys. Rev. A \textbf{56}, 1163 (1997).

\bibitem{s3} C. H. Bennett, G. Brassard, C. Crepeau, R. Jozsa, A. Peres, and
W. K. Wootters, Phys. Rev. Lett. \textbf{70}, 1895 (1993); S. Albeberio
and S. M. Fei, Phys. Lett. A 276, 8 (2000); G. M. D¡¯Ariano,
P. Lo Presti, and M. F. Sacchi, \emph{ibid}. \textbf{272}, 32 (2000); S. Albeverio,
S.-M. Fei, and W.-L. Yang, Phys. Rev. A \textbf{66}, 012301 (2002).

\bibitem{s4} C. H. Bennett and S. J. Wiesner, Phys. Rev. Lett. \textbf{69}, 2881
(1992).

\bibitem{p96} A. Peres, Phys. Rev. Lett. \textbf{77}, 1413 (1996).

\bibitem{ho96} M. Horodecki, P. Horodecki, and R. Horodecki, Phys. Lett. A
\textbf{223}, 1 (1996).

\bibitem{ho97} P. Horodecki, Phys. Lett. A \textbf{232}, 333 (1997).

\bibitem{r1} O. Rudolph, Phys. Rev. A \textbf{67}, 032312 (2003).

\bibitem{r2} K. Chen and L. A. Wu, Quant. Inf. Comput. \textbf{3}, 193 (2003).

\bibitem{r3} M. Horodecki, P. Horodecki, and R. Horodecki, Open Syst. Inf.
Dyn. \textbf{13}, 103 (2006).

\bibitem{r4} K. Chen and L. A. Wu, Phys. Lett. A \textbf{306}, 14 (2002); Phys. Rev.
A \textbf{69}, 022312 (2004); P. Wocjan and M. Horodecki, Open Syst.
Inf. Dyn. \textbf{12}, 331 (2005).

\bibitem{r5} S. Albeverio, K. Chen, and S. M. Fei, Phys. Rev. A \textbf{68}, 062313
(2003).

\bibitem{c1} O. Guhne, P. Hyllus, O. Gittsovich, and J. Eisert, Phys. Rev.
Lett. \textbf{99}, 130504 (2007).

\bibitem{c2} J. D. Vicente, Quant. Inf. Comput. \textbf{7}, 624 (2007).

\bibitem{c3} J. D. Vicente, J. Phys. A: Math. Theor. \textbf{41}, 065309 (2008).

\bibitem{li14} M. Li, J. Wang, S.-M. Fei, and X. Li-Jost, Phys. Rev. A \textbf{89}, 022325
(2014).

\bibitem{Bell} N. Gisin, Phys. Lett. A \textbf{154}, 201 (1991).

\bibitem{yusxia} S. Yu, J.W. Pan, Z.B. Chen and Y.D. Zhang, Phys. Rev. Lett. \textbf{91}, 217903 (2003).

\bibitem{liming}
M. Li and S.M. Fei, Phys. Rev. Lett. \textbf{104}, 240502 (2010).

\bibitem{zmj} M.J. Zhao, T. Ma,  S.M. Fei and Z.X. Wang, Phys. Rev. A \textbf{83}, 052120 (2011).

\bibitem{spe12} C. Spengler, M. Huber, S. Brierley, T. Adaktylos, and B. C. Hiesmayr, Phys. Rev. A \textbf{86}, 022311 (2012).

\bibitem{s5} W. K. Wootters and B. D. Fields, Ann. Phys. (N.Y.) \textbf{191}, 363 (1989).

\bibitem{be} R. A. Bertlmann, K. Durstberger, B. C. Hiesmayr, and P. Krammer, Phys. Rev. A \textbf{72}, 052331 (2005).

\bibitem{s6}S. Brierley and S. Weigert, Phys. Rev. A \textbf{78}, 042312 (2008).

\bibitem{s7} S. Brierley and S. Weigert, Phys. Rev. A \textbf{79}, 052316 (2009).

\bibitem{s8} P. Raynal, X. L$\ddot{u}$, and B.-G. Englert, Phys. Rev. A \textbf{83}, 062303 (2011).

\bibitem{s9} D. McNulty and S. Weigert, J. Phys. A: Math. Theor. \textbf{45}, 102001 (2012).

\bibitem{ka14} A. Kalev and G. Gour, arXiv: 1401.2706v1 [quant-ph] (2014).

\bibitem{qst1} R. B. A. Adamson and A. M. Steinberg, Phys. Rev. Lett. \textbf{105}, 030406 (2010).

\bibitem{qst2} A. Fern$\acute{a}$ndez-P$\acute{e}$rez, A. B. Klimov, and C. Saavedra, Phys. Rev. A \textbf{83}, 052332 (2011).

\bibitem{iv92} I. D. Ivanovic, J. Phys. A: Math. Gen. \textbf{25}, 363 (1992).

\bibitem{weh10} S. Wehner and A. Winter, New J. Phys. \textbf{12}, 025009 (2010).





\end{thebibliography}
\end{document}